\documentclass[conference]{IEEEtran}
\IEEEoverridecommandlockouts
\usepackage{cite}
\usepackage{amsmath,amssymb,amsfonts}
\usepackage{algorithmic}
\usepackage{graphicx}
\usepackage{textcomp}
\usepackage{xcolor}
\usepackage{subfigure}
\usepackage{dblfloatfix}
\def\BibTeX{{\rm B\kern-.05em{\sc i\kern-.025em b}\kern-.08em
    T\kern-.1667em\lower.7ex\hbox{E}\kern-.125emX}}

 \usepackage[normalem]{ulem}
\newcommand{\foothead}[1]{\textcolor{lightgray}{#1 }} 

\usepackage{fancyhdr,lipsum} 

\makeatletter
\def\ps@headings{
\let\@oddhead\@empty
\let\@evenhead\@empty
\let\@oddfoot\@empty
\let\@evenfoot\@empty
}
\def\ps@IEEEtitlepagestyle{
\def\@oddhead{\headerwithDOI}
\let\@evenhead\@empty
\def\@oddfoot{\copyrightnotice}%
\let\@evenfoot\@empty
}
\makeatother

\def\copyrightnotice{
  {
  \begin{minipage}{\textwidth}
  \centering
    \foothead{\copyright2022 IEEE. Personal use of this material is permitted. Permission from IEEE must be obtained for all other uses, in any current or future media, including reprinting/republishing this material for advertising or promotional purposes, creating new collective works, for resale or redistribution to servers or lists, or reuse of any copyrighted component of this work in other works.}
  \end{minipage}
  }
}

\def\headerwithDOI{
  {
  \begin{minipage}{\textwidth}
  \centering
    \foothead{This article has been accepted for publication in the proceedings of the ESSCIRC 2022 - IEEE 48th European Solid State Circuits Conference (ESSCIRC). DOI: 10.1109/ESSCIRC55480.2022.9911527}
  \end{minipage}
  }
}

\begin{document}

\title{Physical Implementation of a Tunable Memristor-based Chua's Circuit}

\author{\IEEEauthorblockN{Manuel Escudero\IEEEauthorrefmark{1}, Sabina Spiga\IEEEauthorrefmark{1}, Mauro di Marco\IEEEauthorrefmark{2}, Mauro Forti\IEEEauthorrefmark{2}, Giacomo Innocenti\IEEEauthorrefmark{3},}
\IEEEauthorblockN{Alberto Tesi\IEEEauthorrefmark{3}, Fernando Corinto\IEEEauthorrefmark{4} and Stefano Brivio\IEEEauthorrefmark{1}}
\IEEEauthorblockA{\IEEEauthorrefmark{1}\textit{CNR—IMM, Unit of Agrate Brianza, Agrate Brianza, 20864, Italy}}
\IEEEauthorblockA{\IEEEauthorrefmark{2}\textit{Università degli Studi di Siena, Siena, 53100, Italy}}
\IEEEauthorblockA{\IEEEauthorrefmark{3}\textit{Università degli Studi di Firenze, Florence, 50121, Italy}}
\IEEEauthorblockA{\IEEEauthorrefmark{4}\textit{Politecnico di Torino, Turin, 10129, Italy}}
}
\maketitle

\begin{abstract}
Nonlinearity is a central feature in demanding computing applications that aim to deal with tasks such as optimization or classification.  Furthermore, the consensus is that nonlinearity should not be only exploited at the algorithm level, but also at the physical level by finding devices that incorporate desired nonlinear features to physically implement energy, area and/or time efficient computing applications. Chaotic oscillators are one type of system powered by nonlinearity, which can be used for computing purposes. In this work we present a physical implementation of a tunable Chua's circuit in which the nonlinear part is based on a nonvolatile memristive device. Device characterization and circuit analysis serve as guidelines to design the circuit and results prove the possibility to tune the circuit oscillatory response by electrically programming the device.
\end{abstract}

\begin{IEEEkeywords}
Memristor, oscillators, chaos, Chua's circuit, nonlinear systems
\end{IEEEkeywords}

\section{Introduction}

The continuous growth in terms of performance and energy efficiency of von-Neumann centralized computer architectures is reaching its limit, which is especially challenging for tasks that deal with huge amounts of data, such as classification or optimization. New paradigms are taking advantage of different forms of analog nonlinear dynamics as an engine for computation, mainly with a two-fold approach: (i) dynamic features operating as working memory, usually referred as local computation or in-memory computing and aiming for massive parallel processing, and (ii) computational performance improvement due to the complexity of nonlinear systems \cite{Kia2017}.

The focus of this work lies on the Chua's circuit, an autonomous oscillator that exhibits a plethora of behaviors, including chaotic oscillations \cite{Kennedy1993}, which can be used as the basis for novel nonlinear computing systems, like oscillator-based computing \cite{Csaba2020} and reservoir computing \cite{Tanaka2019}. For this circuit to show such complex oscillations, a nonlinear element is required, which usually is built using operational amplifiers or diodes \cite{Fortuna2009}. However, these solutions do not offer configurability and technological scalability at the same time.

The possibility of exploiting memristor nonlinearity of their I-V characteristics in the Chua's circuit has already been considered, e.g. \cite{Itoh2008}. Memristors are two terminal devices with nonlinear conduction mechanisms holding an internal state, e.g. its resistance, that depends on its previous history. Memristive devices are attractive due to their highly scalable, non-volatile memory properties and different dynamical features they can provide. However, to our best knowledge, there has yet not provided experimental proof of a Chua's circuit implementation using real memristive devices.  

In this paper, a physical hardware implementation of a memristor-based Chua's circuit is demonstrated. Thanks to the tunable nonlinearity of the I-V characteristics, we demonstrate bifurcations and chaos, observed from experimental circuit responses. Our results are achieved by carefully matching both the requirements of the circuit and of the device characteristics.

\section{Tunable Chua's Circuit enabled by memristor nonlinearity} \label{memChuaCircuit}

\subsection{Memristor characteristics}

\begin{figure}[!t]
\centerline{\includegraphics[width=0.98\linewidth]{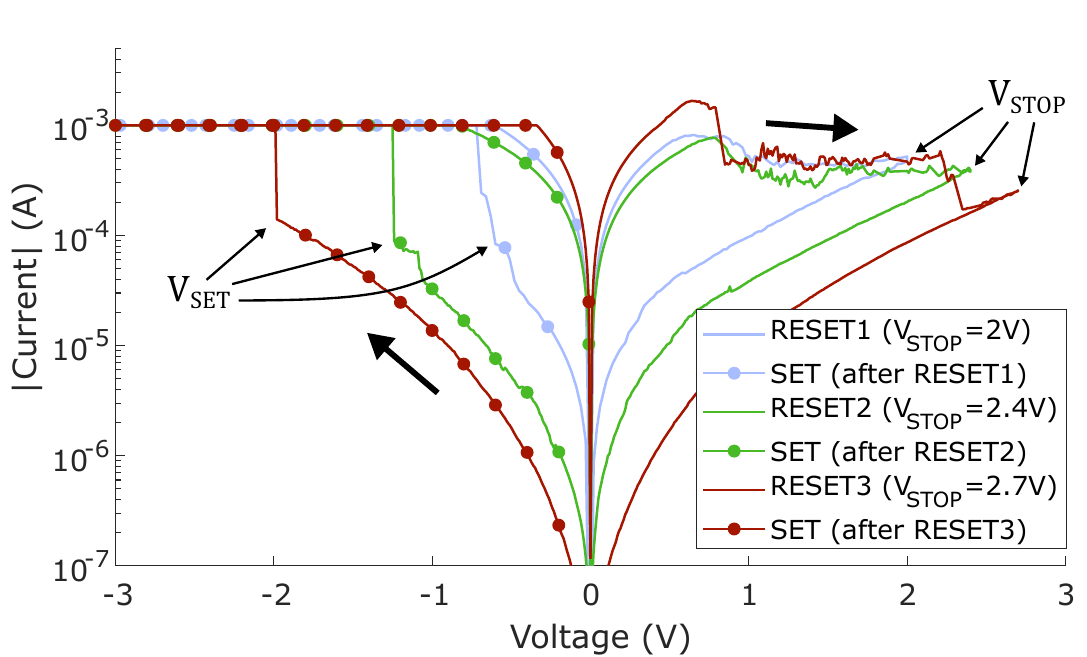}}
\caption{Pt/HfO$_2$/TiN device switching operation, displaying three RESET/SET cycles using $V_{STOP}$ = 2, 2.4 and 2.7 V.}
\label{fig:deviceOperation}
\end{figure}

The nonvolatile memristive devices used in this work are based on a Pt/HfO$_2$/TiN stack prepared as already discussed in \cite{Brivio2015}. The devices are measured through a B1500 Keysight semiconductor parameter analyzer. Fig. \ref{fig:deviceOperation} reports a representative switching operation. A high negative voltage produces the creation of a first filament in a fresh device (not shown) \cite{Brivio2015}, which initiates the bipolar switching with RESET (SET) operations at positive (negative) voltages. A compliance current at 1 mA is required to prevent devices breakdown. RESET operation partially dissolves the formed filament (high resistance state, HRS) and SET operations reinstate it (low resistance state, LRS) under a compliance current control. The resistance states are retained for more than 10 years and can be programmed for thousands of cycles \cite{Brivio2017} \cite{Frascaroli2015}. The HRS resistance value can be controlled by the maximum voltage applied during the RESET sweep (V$_{STOP}$). We notice that the higher the HRS resistance the higher the SET transition voltage (V$_{SET}$, vertical transition) in the negative polarity.

The HRS current-voltage ($i_M-v_M$) characteristics is nonlinear and is well described by a 5$^{th}$ order polynomial from V$_{SET}$ to V$_{STOP}$, as
\begin{equation}
i_M=p_1v_M + p_2v_M^2+ ... +p_5v_M^5   \label{polynomialModel}
\end{equation}

\subsection{Proposed Chua's circuit}

\begin{figure*}[!t]
\centering
\subfigure[]{\includegraphics[width=2.2in]{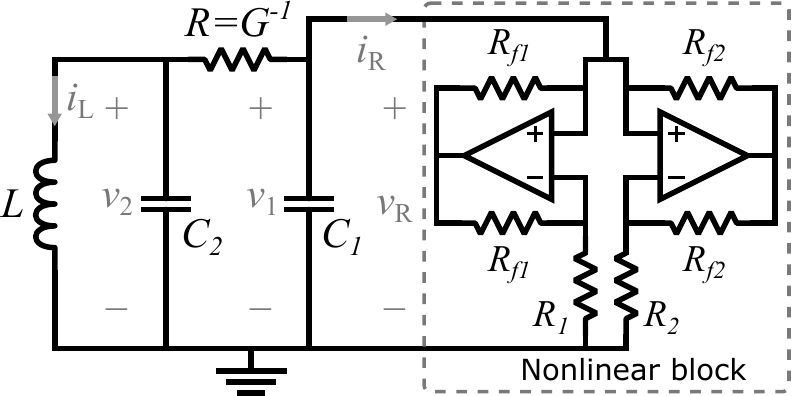} 
\label{fig:chuaScheme}}
\hfil
\subfigure[]{\includegraphics[width=2.2in]{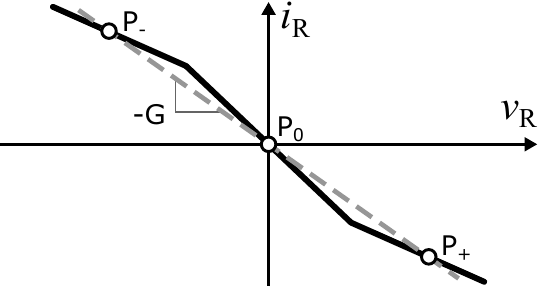} \label{fig:chuaDiode}}
\hfil
\subfigure[]{\includegraphics[width=2.2in]{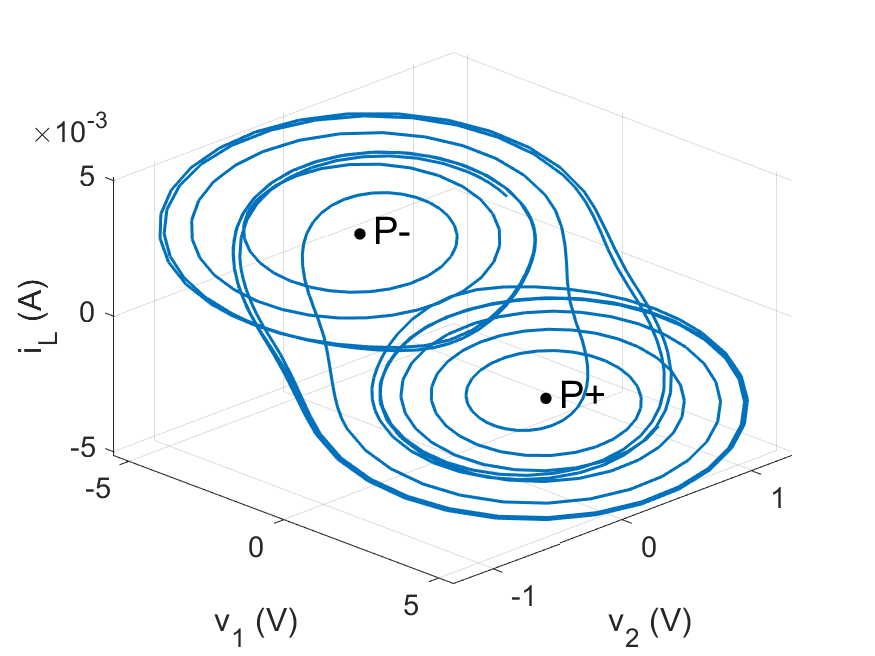} \label{fig:chuaTraj}}
\\
\subfigure[]{\includegraphics[width=2.2in]{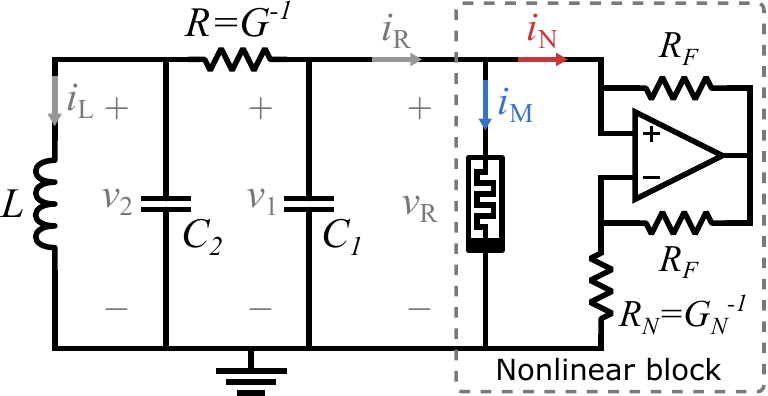}
\label{fig:memChuaScheme}}
\hfil
\subfigure[]{\includegraphics[width=2.2in]{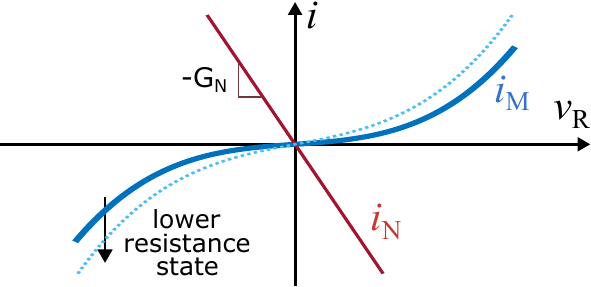}
\label{fig:IVmemristorNIC}}
\hfil
\subfigure[]{\includegraphics[width=2.2in]{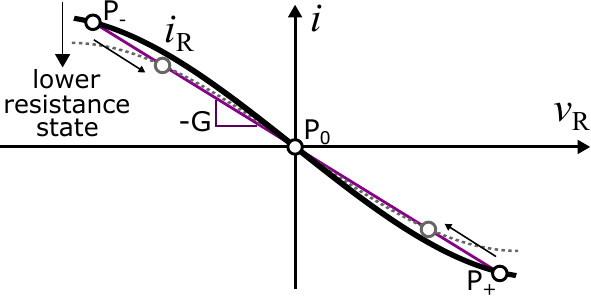}
\label{fig:IVsum}}
\caption{Original Chua's circuit (a) scheme, (b) detailed nonlinear block i-v characteristic and (c) example of a trajectory of a double-scroll attractor in the state space. Proposed memristor-based Chua's circuit (d) scheme, (e) memristor i-v characteristics at two different HRS extracted from measurement data (in blue) and the negative impedance converter (in red), and (f) complete nonlinear block i-v characteristic for both resistance states.}
\label{fig:chuaCircuit}
\end{figure*}

Fig. \ref{fig:chuaScheme} depicts the original Chua's circuit \cite{Kennedy1993}, containing different linear passive elements as well as the nonlinear block with a piecewise linear voltage-dependent current $i_R$ of Fig. \ref{fig:chuaDiode}. The state of this circuit is defined by the three state variables $v_1$, $v_2$ and $i_L$. Three equilibrium points exist, labelled as $P_0$, $P_+$ and $P_-$, which for the variable $v_1$ are defined as solutions of
\begin{equation}
i_R(v_1) + G v_1 = 0 \label{eqPoints}
\end{equation}
or graphically found at the crossing of $i_R$ with the load line $-G$, as shown in \ref{fig:chuaDiode}. Note that given such $i_R$, $P_+$ and $P_-$ only exist if
\begin{equation}
di_R/dv(0) < -G \label{localNegConductance}
\end{equation}
where $di_R/dv(0)$ is the nonlinear block conductance at $P_0$.

The circuit shows a variety of trajectories, i.e. the evolution of the state variables through time, on the basis of the values of the circuit parameters and stability of the equilibrium points: periodic and nonperiodic oscillations around $P_+$ or $P_-$, or nonperiodic oscillations around both $P_+$ and $P_-$ (double-scroll, shown in Fig. \ref{fig:chuaTraj}).

The memristor-based Chua's circuit in our study is shown in Fig. \ref{fig:memChuaScheme}, where the only difference is found in the nonlinear block. This block is composed of two elements connected in parallel: the nonvolatile device, from now on referred to as memristor, and an operational amplifier implementing a negative resistor $-R_N$. These elements contribute with a current $i_M$ from \eqref{polynomialModel} and $i_N = -G_Nv_R$, as depicted in Fig. \ref{fig:IVmemristorNIC}. Summing both currents in the parallel association, the total current $i_R$ is depicted in Fig. \ref{fig:IVsum}. Smooth nonlinear functions with similar characteristics to $i_R$ have been already confirmed to show chaotic attractors when used in the Chua's circuit \cite{KHIBNIK1993}.

The complete memristor-based Chua's circuit can be described as a dynamical system using the state equations
\begin{equation}
  \begin{array}{@{}l@{}}
    \frac{dv_1}{dt}=\frac{1}{C_1}\left( (v_2 - v_1)G - i_R(v_1)  \right)  \\
    \frac{dv_2}{dt}=\frac{1}{C_2}\left( (v_1 - v_2)G + i_L  \right)  \\
    \frac{di_L}{dt}=-\frac{v_2}{L}  \\    
  \end{array} \label{stateEq}
\end{equation}

where the different parameters are reported in Fig. \ref{fig:memChuaScheme}. As in the original Chua's circuit, three equilibrium points exist given $i_R$ shown in \ref{fig:IVsum} if \eqref{localNegConductance} is satisfied (in this case $di_R/dv(0)=p_1-G_N$); defined by \eqref{eqPoints}, $P_0$ is at $v_1$ = 0 V, while $P_+$ and $P_-$ depend on the memristor i-v characteristics, $G$ and $G_N$. The circuit trajectory is expected to be similar to the one in Fig. \ref{fig:chuaTraj}. It is worth noting that this implementation does not only benefit from the memristor nonlinear characteristics but the nonlinear block can be electrically tuned by programming the memristor to different HRS, as suggested in Fig. \ref{fig:IVmemristorNIC} and \ref{fig:IVsum}. In this case, it is expected that if the memristor resistance state is decreased, $P_+$ and $P_-$ shift towards $P_0$, as shown in Fig. \ref{fig:IVsum}.

\section{Circuit design considering physical memristor constraints} \label{circuitDesign}

In this section, we present a design approach that matches device features and circuit requirements.

\begin{figure}[!t]
\centering
\subfigure[]{\includegraphics[width=0.98\linewidth]{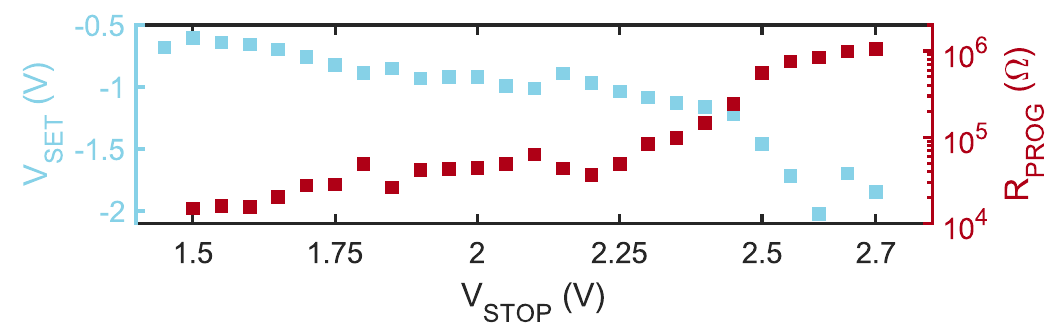} \label{fig:Vset}}
\\
\subfigure[]{\includegraphics[width=0.95\linewidth]{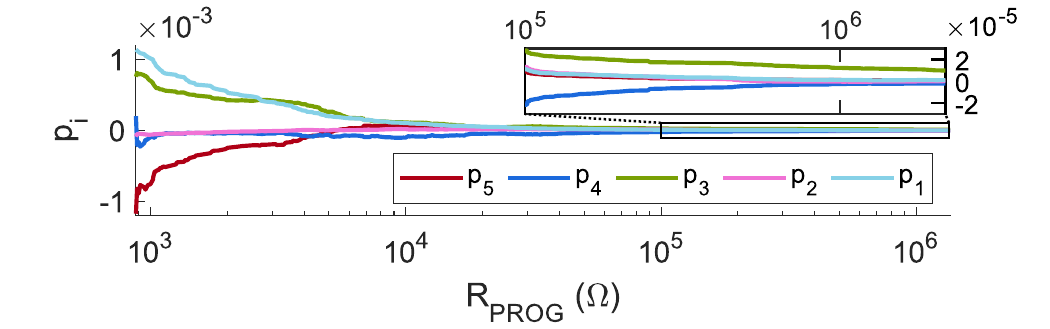} \label{fig:Param}}
\caption{(a) Average $V_{SET}$ and resistance measured at 0.1 V $R_{PROG}$ after performing a progressive reset operation according to different used $V_{STOP}$. (b) Average fitting parameters for different $R_{PROG}$.}
\label{fig:VsetandP}
\end{figure}

The circuit needs a static nonlinear characteristics, that the memristor can provide as long as $v_1$ (the voltage across the memristor) is not extended below $-V_{SET}$ or above $V_{STOP}$. From Fig. \ref{fig:deviceOperation} it is clear that $|V_{SET}|<V_{STOP}$, thus $V_{SET}$ is the most limiting value. For this reason, $V_{SET}$ has been characterized by programming a single device using a progressive RESET procedure, in which successive RESET operations increasing the voltage are applied until reaching a target $V_{STOP}$. After each progressive RESET, a SET operation is carried out to capture $V_{SET}$. A single device is programmed with progressive RESET with $V_{STOP}$ values from 1.5 to 2.7 V and repeated 10 times to account for the intrinsic memristor cycle-to-cycle variability. Results in fig. \ref{fig:Vset} show how  $|V_{SET}|$ generally increases with the used $V_{STOP}$. Moreover, the memristor resistance state increases with $V_{STOP}$, where $R_{PROG}$ is the memristor resistance measured at 0.1 V after applying the progressive RESET. Each programmed state is fitted with \eqref{polynomialModel} and the fitting parameters $p_i$ are depicted in \ref{fig:Param} as a function of $R_{PROG}$. It is worth noting that the linear term parameter $p_1$ is a good estimator of the memristor conductance at low voltage. As $R_{PROG}$ increases, $p_1$ becomes comparable with the nonlinear terms parameters. As a result, it is convenient to program the device to higher resistance states to achieve both high $|V_{SET}|$ and nonlinear characteristic.


Next, the circuit is designed to ensure operation inside the voltage range supported by the memristor and at the same time to exhibit oscillatory behavior. Specifically, we look for the conditions to reproduce a double-scroll attractor analogous to the scenario presented in \cite{Kennedy1993}. Oscillations are guaranteed as long as the three equilibrium points in the system are unstable. At $P_0$, the instability is guaranteed if the trace of the Jacobian matrix of \eqref{stateEq} evaluated at this point is forced to zero, which occurs only if

\begin{equation}
G_N = G + \frac{C_1}{C_2} G + p_1   \label{GN}
\end{equation}
Note that \eqref{GN} implies \eqref{localNegConductance}. Additionally, it can be easily observed that $P_+$ and $P_-$ are also unstable when \eqref{GN} is used, again by analyzing the Jacobian matrix trace evaluated at $P_+$ and $P_-$. In conclusion, $G_N$ calculated from \eqref{GN} guarantees the three equilibrium points existence and their instability.

Now, by combining \eqref{eqPoints} and \eqref{GN}, \begin{equation}
G = \frac{C_2}{C_1} \left[\frac{i_R(v_{eq})}{v_{eq}}-p_1 \right]   \label{G}
\end{equation}
where $v_{eq}$ is the absolute value of the $v_1$ coordinate for $P_+$ and $P_-$. This last expression is particularly useful, because it allows to design $G$ (and $G_N$) by fixing $P_+$ and $P_-$ position, which are in turn related with the amplitude of the trajectory in $v_1$ dimension.

Finally, $C_1$, $C_2$ and $L$ values influence the attractor exhibited by the circuit. Guided by the bifurcation analysis in \cite{Kennedy1993}, a double-scroll attractor can be achieved by fixing the adimensional parameters of Chua's circuit to $\alpha$ = 10 and $\beta$ = 14.22, where
\begin{equation}
\alpha = \frac{C_2}{C_1} \quad\mathrm{,}\quad \beta = \frac{R^2C_2}{L}  \label{alphaBeta}
\end{equation}

Following the previous device physical constraints and previous exposed conditions, the Chua's circuit has been designed by programming a memristor with $V_{STOP}$ = 2.6 V. More especifically, component values $C_2$, $L$, $G$ and $G_N$ are calculated with \eqref{GN}-\eqref{alphaBeta} using the fitting parameters of the programmed memristor i-v characteristics, fixing $P_+$ and $P_-$ at $v_{eq} = 0.9$ V (lower than $|V_{SET}|$ at $V_{STOP}$ = 2.6 V, as shown in Fig. \ref{fig:Vset}) and choosing $C_1$ to 10 nF. Both fitting parameters and circuit impedances values used are shown in Table \ref{tab1}.

\begin{table}[!t]
\caption{Table I. Memristor I-V characteristics fitting parameters ($V_{STOP}=2.6$ V) used in the design and impedances calculated}
\begin{center}
\begin{tabular}{cccccc}
\hline
\textbf{$p_1$}& 1.91$\cdot 10^{-6}$ & \textbf{$p_2$}& 3.11$\cdot 10^{-7}$ & \textbf{$p_3$}& 1.91$\cdot 10^{-5}$ \\
\hline
\textbf{$p_4$}& -5.20$\cdot 10^{-6}$ &
\textbf{$p_5$}& 1.77$\cdot 10^{-6}$ &
\textbf{$v_{eq}$} & 0.9 V \\
\hline
\textbf{$R$} & 7.643 k$\Omega$ & \textbf{$R_{N}$} & 6.856 k$\Omega$ & \textbf{$L$} & 410 mH \\
\hline
\textbf{$C_1$}& 10 nF &
\textbf{$C_2$}& 100 nF &
  &   \\
\hline
\end{tabular}
\label{tab1}
\end{center}
\end{table}

\begin{figure*}[!t]
\centerline{\subfigure[]{\includegraphics[width=2.2in]{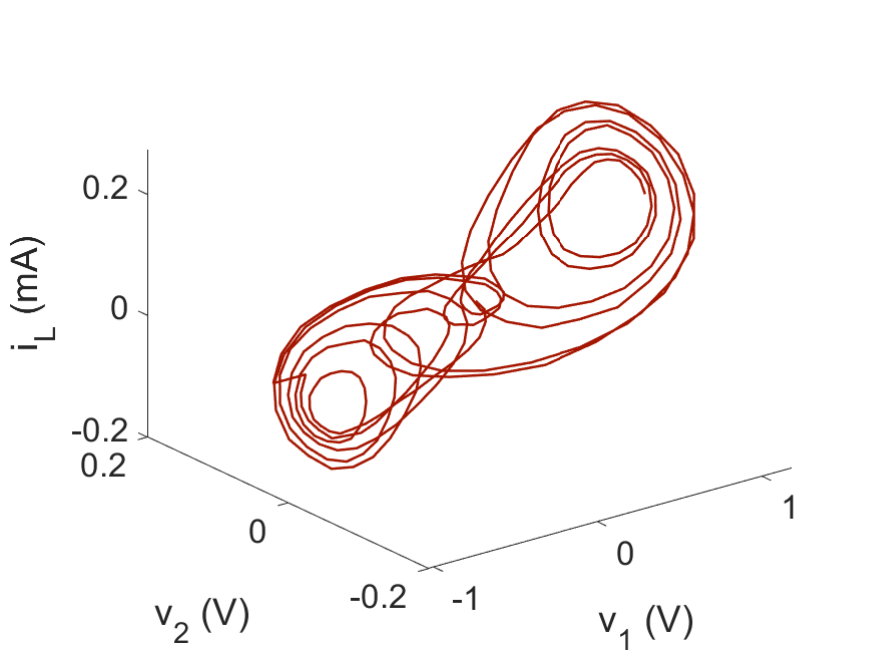} 
\label{fig:designTrajectory}}
\hfil
\subfigure[]{\includegraphics[width=2.2in]{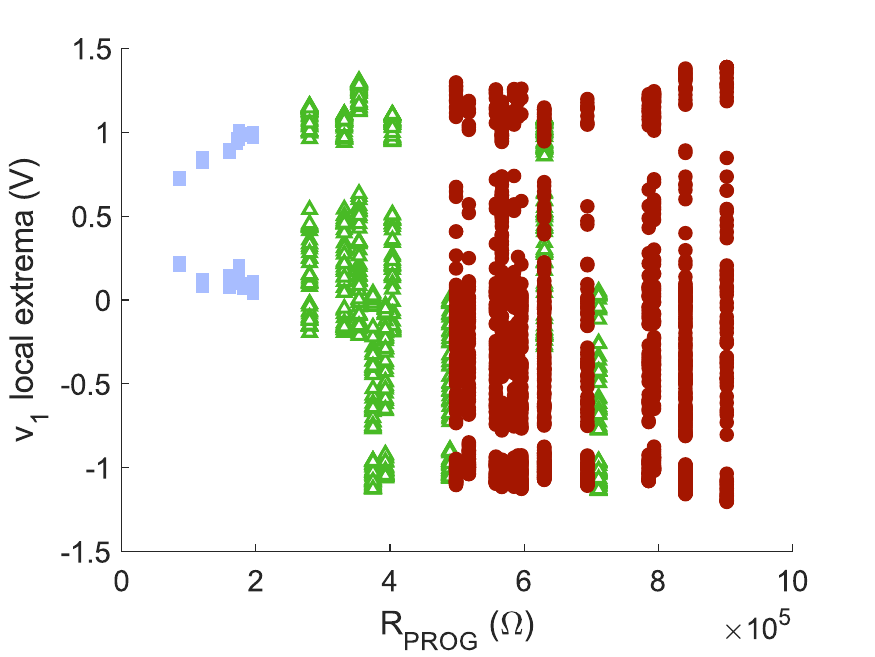} \label{fig:tunability2}}
\hfil
\subfigure[]{\includegraphics[width=2.2in]{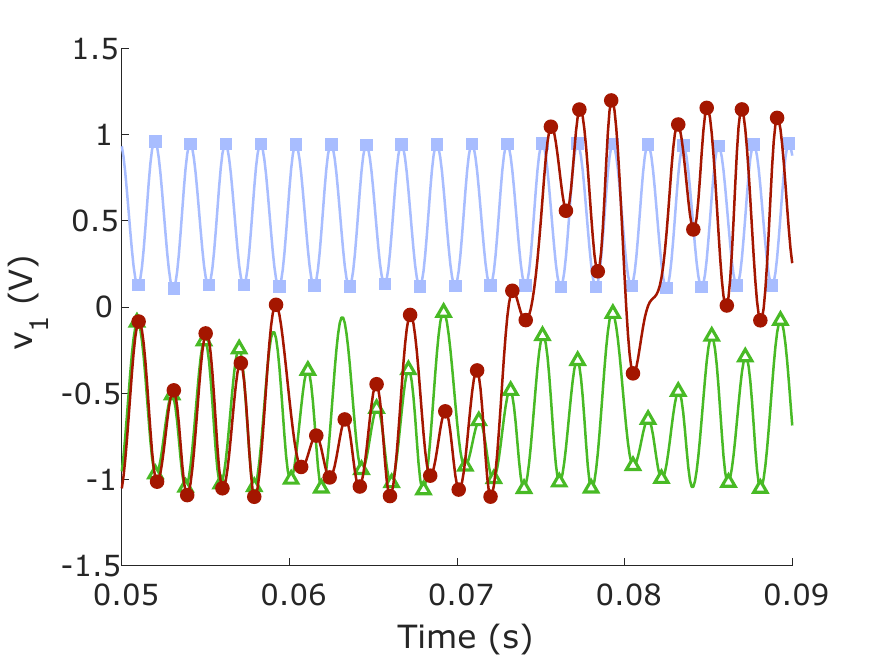} 
\label{fig:tunability1}}
}
\caption{Experimental validation results of the tunable memristor-based Chua's circuit. (a) Trajectory obtained after designing the circuit to show a double-scroll attractor. (b) All trajectories obtained after each programming operation; only the local extrema of $v_1$ are shown, coloured according to the types of identified trajectories: double-scroll attractor in red, single-scroll attractor in green and periodical oscillation in blue. (c) Examples of the $v_1$ temporal evolution for each kind of identified trajectory.}
\label{fig:tunabilityResults}
\end{figure*}

\section{Physical implementation and experimental validation} \label{results}

The previously designed Chua's circuit has been built, connected to the memristor and the state variables are captured. Fig. \ref{fig:designTrajectory} shows the resulting trajectory, which is indeed the double-scroll attractor, with $P+$ and $P-$ located approximately at $v_1 = \pm$0.9 V, validating our design procedure. Notice the asymmetry of the trajectory, probably due an slight asymmetry in the memristor i-v characteristics or non-idealities of the circuit elements.

Fig. \ref{fig:tunability2} reports an experimental bifurcation plot as a function of the $R_{PROG}$ demonstrating that the memristor is properly providing the nonlinearity required for chaotic behaviour and that it allows tuning the dynamics from periodic trajectory (blue squares), single-scroll attractor (green triangles) to double-scroll attractor (red circles). The bifurcation plot only depicts the local extrema of $v_1$, which tells us qualitatively the type of obtained oscillation. For a better understanding of the results, an example of each kind of oscillatory trajectory is also included in Fig. \ref{fig:tunability1}. Note that both the single-scroll attractor and the periodic trajectories oscillate either around $P_+$ or $P_-$, which depends only on the initial circuit conditions, not controlled during our experiments. 

It is worth mentioning that, generally, the voltage spanning of $v_1$, which is the voltage drop on the memristor, is wider for higher $R_{PROG}$ in agreement with the displacement of the equilibrium points $P_+$ and $P_-$ depicted in Fig. \ref{fig:IVsum}. Such growth of $v_1$ spanning range with $R_{PROG}$ does not lead to unintentional memristor state perturbation as also $V_{SET}$ grows with $R_{PROG}$. Such combination helps the circuit design.

Although the bifurcation plot in Fig. \ref{fig:tunability2} shows that for some $R_{PROG}$ bands it is expected to obtain a certain type of trajectory, we have found few single-scroll attractor among resistance states in which double-scroll attractors are dominant. The same observation is valid for the general trend of decreasing amplitude with $R_{PROG}$. The main cause is the the device intrinsic cycle-to-cycle variability, that hinders repeatable trajectories.

\section{Conclusions} \label{conclusions}

We successfully implemented physically a Chua's circuit powered by the nonlinearity of the nonvolatile Pt/HfO$_2$/TiN device. Moreover, we showed that circuit tunability is possible using nonvolatile resistive switching devices, and different kinds of trajectories can be obtained after changing the memristor state. All these achievements have been possible after a careful design based on the dynamic system analysis and the constraints observed from the device characterization. By using a dedicated nonlinear programmable device for this circuit, the circuit complexity is alleviated and configurability features are provided at the same time. This offers an attractive alternative for implementing scalable chaotic oscillators. For real applications, the programming operation may need to control precisely the memristor state in order to guarantee a robust tuning. Therefore, program and verify techniques can be considered for this purpose.

\section*{Acknowledgment}

This work is partially supported by the PRIN-MIUR project COSMO (Prot. 2017LSCR4K).

\bibliography{bib/IEEEabrv.bib,bib/ESSCIRC22.bib}{}
\bibliographystyle{IEEEtran}

\end{document}